\documentclass[aps,prl,twocolumn,showpacs,superscriptaddress]{revtex4}
\usepackage{graphicx}

\begin{document}

\title{Quantum Critical Scaling
and origin of Non-Fermi-Liquid behavior in
Sc$_{1-x}$U$_{x}$Pd$_3$}

\author{Stephen D. Wilson}
\affiliation{ Department of Physics and Astronomy, The University
of Tennessee, Knoxville, Tennessee 37996-1200, USA }
\author{Pengcheng Dai}
\email{daip@ornl.gov} \affiliation{ Department of Physics and
Astronomy, The University of Tennessee, Knoxville, Tennessee
37996-1200, USA } \affiliation{ Condensed Matter Sciences
Division, Oak Ridge National Laboratory, Oak Ridge, Tennessee
37831, USA}
\author{D. T. Adroja}
\affiliation{ ISIS Facility, Rutherford Appleton Laboratory,
Didcot, Oxon OX11 0QX, United Kingdom}
\author{S.-H. Lee}
\affiliation{ NIST Center for Neutron Research, National Institute
of Standards and Technology, Gaithersburg, Maryland 20899, USA}
\author{J.-H. Chung}
\affiliation{ NIST Center for Neutron Research, National Institute
of Standards and Technology, Gaithersburg, Maryland 20899, USA}
\affiliation{ Department of Materials Science and Engineering,
University of Maryland, College Park, Maryland 20742, USA }
\author{J. W. Lynn}
\affiliation{ NIST Center for Neutron Research, National Institute
of Standards and Technology, Gaithersburg, Maryland 20899, USA}
\author{N. P. Butch}
\affiliation{ Department of Physics and Institute for Pure and
Applied Physical Sciences, University of California, San Diego, La
Jolla, California 92093-0319, USA}
\author{M. B. Maple}
\affiliation{ Department of Physics and Institute for Pure and
Applied Physical Sciences, University of California, San Diego, La
Jolla, California 92093-0319, USA}
\begin{abstract}
We use inelastic neutron scattering to study magnetic excitations
of Sc$_{1-x}$U$_{x}$Pd$_3$ for U concentrations ($x=0.25$,
0.35) near the spin glass quantum critical point (QCP). The
excitations are spatially incoherent, broad in energy
($E=\hbar\omega$), and follow $\omega/T$ scaling at all
wave vectors investigated. Since similar $\omega/T$ scaling has
been observed for UCu$_{5-x}$Pd$_x$ and CeCu$_{6-x}$Au$_{x}$ near
the antiferromagnetic (AF) QCP, we argue that the observed non
Fermi liquid (NFL) behavior in these $f$-electron materials arises
from the critical phenomena near a $T=0$ K phase transition,
irrespective of the nature of the transition.
\end{abstract}


\maketitle


The breakdown of Fermi-liquid theory has been observed in a class of strongly
correlated $f$-electron materials, following the original discovery of
this so-called non-Fermi-liquid (NFL) behavior in the
Y$_{1-x}$U$_x$Pd$_3$ pseudobinary alloy in 1991
\cite{Seaman,Andraka,Maple,Stewart}. In spite of intensive
theoretical and experimental efforts over the past decade
\cite{Cox,Affleck,Sachdev,Sengupta,Georges,Si,Miranda,Neto,Grempel},
it is still unclear whether the observed NFL behavior is an
intrinsic property \cite{Gajewski,Dickey} or extrinsic property
associated with metallurgical inhomogeneity in these materials
\cite{Sullow}. Models describing this anomalous NFL behavior
include single-ion physics of noninteracting local magnetic
moments \cite{Cox,Affleck}, close proximity to a $T=0$ K
second-order phase transition or quantum critical point (QCP)
\cite{Sachdev,Sengupta,Georges,Si}, and disorder induced effects
\cite{Miranda,Neto,Grempel}.

We studied Sc$_{1-x}$U$_{x}$Pd$_3$ because this system has a phase
diagram and NFL properties similar to Y$_{1-x}$U$_x$Pd$_3$, but with a nearly
homogeneous U distribution in the ScPd$_3$ matrix \cite{Gajewski}.
While the U inhomogeneity in Y$_{1-x}$U$_x$Pd$_3$ is unlikely the
main cause of the NFL behavior \cite{Dickey}, neutron scattering
experiments seeking to provide constraints on various microscopic
models have reached different conclusions. According to Lea,
Leask, and Wolf \cite{lea}, the cubic crystalline electric field
(CEF) of ScPd$_3$ splits the U$^{4+}$ $J=4$ multiplet into
$\Gamma_4$ and $\Gamma_5$ triplets, a $\Gamma_1$ singlet, and a
$\Gamma_3$ doublet. If the single-ion based two-channel
quadrupolar Kondo effect (QKE) is responsible for the NFL behavior
in Y$_{0.8}$U$_{0.2}$Pd$_3$ \cite{Seaman}, the U$^{4+}$ ground
state should be a nonmagnetic $\Gamma_3$ with magnetic $\Gamma_5$
and $\Gamma_4$ excited states \cite{Cox}. In contrast, polarized
triple-axis neutron-scattering experiments on
Y$_{1-x}$U$_{x}$Pd$_3$ reveal a magnetic ground state for $x=0.45$
and possibly for $x=0.2$, thus precluding the possibility of a QKE
\cite{Dai}. However, based on subsequent neutron time-of-flight
measurements, Bull {\it et al.} \cite{Bull} argue that the
$\Gamma_3$ doublet ground state is more consistent with the
$x=0.45$ data and the $x=0.2$ compound has a degenerate $\Gamma_3$
and $\Gamma_5$ ground state. In this case, the QKE could be the
predominant cause of NFL behavior \cite{Dickey}.

In this Letter, we report neutron scattering
experiments on Sc$_{1-x}$U$_{x}$Pd$_3$ ($x=0.0$, $0.25$, $0.35$).
We show that magnetic excitations at the NFL concentration
($x=0.35$) do not form the distinct CEF excitations seen in
Y$_{0.55}$U$_{0.45}$Pd$_3$. Instead, the susceptibility
$\chi^{\prime\prime}(q,\omega,T)$ at all probed wave vectors
($q$), temperatures ($T$), and energies ($\hbar\omega$) obeys
$\omega/T$ scaling indicative of a $T = 0$ K second order phase
transition. While such behavior is also observed in the NFL
compounds CeCu$_{6-x}$Au$_x$ \cite{Schroeder}, UCu$_{5-x}$Pd$_x$ \cite{Aronson,Aronson2}, and Ce(Rh$_{0.8}$Pd$_{0.2}$)Sb \cite{park} near antiferromagnetic
(AF) QCP, $\chi^{\prime\prime}(q,\omega,T)$ in
Sc$_{1-x}$U$_{x}$Pd$_3$ ($x=0.35$) is wave vector independent with no spatial correlations and obeys $\omega/T$ scaling over a much wider energy range with a different critical exponent. Therefore, the dynamics of isolated U
ions are responsible for the temperature and energy scaling,
suggesting that the NFL behavior originates from the spin-glass
phase transition suppressed to near zero temperature.

\begin{figure}[t]
\includegraphics[scale=.3]{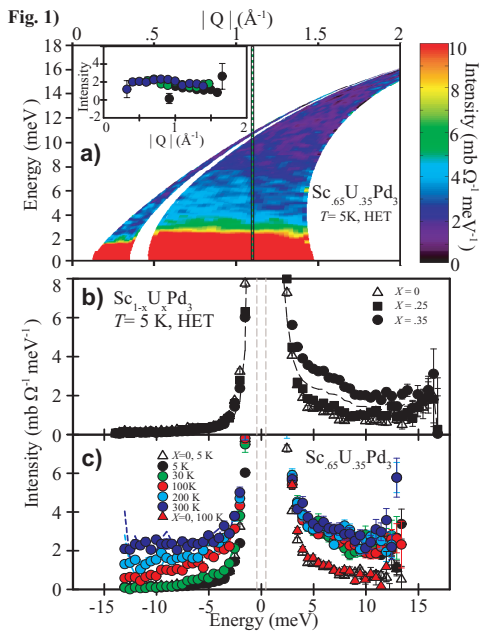}
\caption{(a) $E$-$q$ image of Sc$_{1-x}$U$_{x}$Pd$_3$ ($x=0.35$)
at 5 K for $E_i=18$ meV. The inset shows $E$-integrated [$E=3$-6 meV (blue); 6-9 meV (green); 9-12 meV (black)] $q$-cuts at 5 K. (b) The $q$-integrated
($0<q<2.5$ \AA$^{-1}$) $S(\omega,T)$ for $x=0.0$, 0.25, and 0.35.
The dashed line shows the expected magnetic scattering for
$x=0.25$ assuming simple U-concentration scaling. (c)The
$q$-integrated ($0<q<1.5$ \AA$^{-1}$) $S(\omega,T)$ for $x=0.35$ at
various temperatures. Dashed lines on the $E<0$ side
reflect detailed balance expectations calculated from the magnetic scattering on the $E>0$ side. Grey,
vertical dashed-lines show the resolution half-width of 0.433 meV.}
\end{figure}

Our experiments were performed on the HET time-of-flight
spectrometer at the UK ISIS spallation neutron source \cite{Bull},
and on the BT-2 and cold neutron SPINS triple-axis spectrometers at the NIST
Center for Neutron Research (NCNR). The HET data were collected
with incident beam energies ($E_i$) of both 18 meV and 65 meV for
a range of temperatures, and a vanadium standard was used to
normalize the scattering intensity to absolute units. The magnetic
scattering in U-doped materials was determined by comparing the
scattering intensity with that of the nonmagnetic ScPd$_3$ parent
compound for 18 meV data, and by subtraction of a parent-compound-generated mapping background for 65 meV data \cite{Bull,Murani}.
To study the low-energy spin dynamics, we used SPINS with final neutron energy fixed at
$E_{f} = 5$ meV. An incident beam collimation of $80^\prime$ was
used followed by a cold Be filter and a radial collimator after
the sample. We also collected data using polarized neutrons on
BT-2 to separate the magnetic signal from nuclear spin incoherent
scattering. For the experiment, we prepared 18 g polycrystalline
samples of Sc$_{1-x}$U$_{x}$Pd$_3$ ($x=0$, 0.25, 0.35) through
arc-melting techniques \cite{Gajewski}. The lattice parameters of
these cubic Cu$_3$Au structure materials are $a = b = c = 4.01$
\AA\ for $x=0.35$ and 3.99 \AA\ for $x=0.25$.

\begin{figure}[t]
\includegraphics[scale=.3]{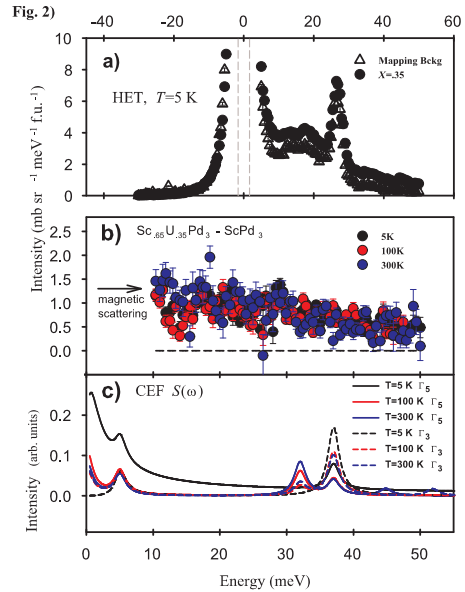}
\caption{(a) $q$-integrated ($0<q<2.5$ \AA$^{-1}$) $S(\omega,T)$
for $x=0.35$ with $E_i$=65 meV. Open triangles show the mapping
background, and vertical dashed lines are the resolution half-width
of 1.65 meV. (b) Net magnetic scattering at various temperatures.
(c) Calculated CEF $S(\omega,T)$ for two possible ground states:
magnetic $\Gamma_5$ and nonmagnetic $\Gamma_3$.}
\end{figure}

Figure 1 summarizes HET results with $E_i=18$ meV for
Sc$_{1-x}$U$_{x}$Pd$_3$ at $x=0$, $0.25$,
0.35, where we probed excitations in the energy range between 3
and 13 meV. The scattering for $x=0.35$ in the energy-momentum
($E$-$q$) space probed (Fig. 1a) shows a broad continuum of
intensity with no peak at the expected AF ordering wave vector for
Y$_{0.55}$U$_{0.45}$Pd$_3$ marked as the vertical dashed line
\cite{Dai}. Different energy-integrated cuts at 5 K (see inset of
Fig. 1a) show no modulation at any wavevector, different from that of Y$_{1-x}$U$_{x}$Pd$_3$ (Fig. 10 of Ref. \cite{Bull}). To see if the scattering in Fig.
1a is magnetic, we compare $q$-integrated energy cuts for
all three concentrations at 5 K (Fig. 1b). While the
outcome shows clear magnetic response for the two doped systems,
the scattering is broad and featureless with no evidence for
localized CEF states. In addition, the magnetic scattering does not
follow the U concentration scaling. Assuming that the magnetic
fluctuations in Sc$_{1-x}$U$_{x}$Pd$_3$ scale linearly with the U
solute concentration, one would expect scattering for $x=0.25$ as
the dashed line in Fig. 1b. The actual scattering from the
$x=0.25$ concentration instead almost lies directly on top of the
nonmagnetic parent background with much less magnetic signal. We
note that similar behavior has also been observed in
Y$_{1-x}$U$_{x}$Pd$_3$ \cite{Bull}.

Since the Sc$_{1-x}$U$_{x}$Pd$_3$ $x=0.25$ compound is nearly
nonmagnetic and does not exhibit strong NFL features
\cite{Dickey}, we focus on the NFL $x=0.35$ compound and study the
temperature evolution of the magnetic scattering. The
most striking feature of the data is the temperature independence
on the neutron energy loss side of the spectra, while the neutron
energy gain side obeys detailed balance as shown in Fig. 1c. To confirm that such behavior
indeed arises from the U magnetic moment, we performed a careful
study of the temperature dependence of the nonmagnetic ScPd$_3$ and found that the nonmagnetic scattering is
temperature independent below 100 K and increases only slightly at
300 K (Fig. 1c).

\begin{figure}[t]
\includegraphics[scale=.3]{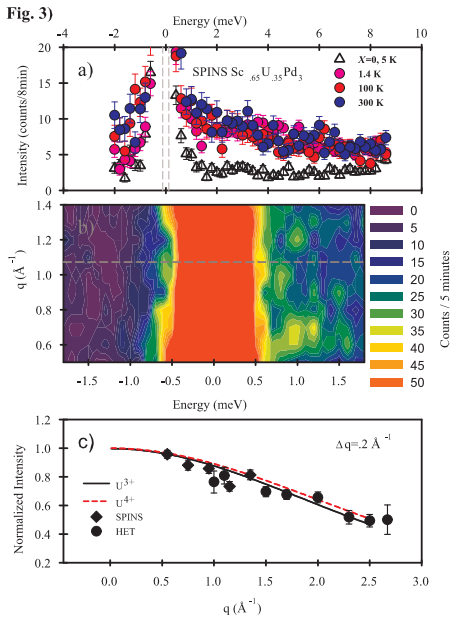}
\caption{(a) Energy scans at fixed $q=1.3$ \AA$^{-1}$ for $x=0.35$
and 0.0 on SPINS. Dashed vertical lines show the resolution
halfwidth of $0.12$ meV. (b) $(q,E)$ map of magnetic fluctuations
for $x=0.35$ at 1.4 K. The dashed grey line shows the AF ordering
wave vector of $(0.5,0.5,0)$ for Y$_{0.55}$U$_{0.45}$Pd$_3$
\cite{Dai}. (c) $q$-dependence of magnetic excitations. Solid
lines show calculated magnetic form factors for U$^{3+}$ and
U$^{4+}$ ions.}
\end{figure}

To determine the magnetic excitations of the $x=0.35$ compound
above 13 meV, we increased $E_i= 65$ meV at the HET. Fig. 2a shows the scattering at $T= 5$ K for
both the $x=0.35$ compound and the nonmagnetic background
\cite{Bull,Murani}. The resulting difference spectra at several
temperatures are shown in Fig. 2b. Similar to Fig. 1c, the
magnetic excitations are broad, temperature independent, and
extend up to 50 meV. If excitations from the U moments in the $x=0.35$
compound have localized states at $\sim$6 meV and $\sim$36 meV as
in the AF ordered Y$_{0.55}$U$_{0.45}$Pd$_3$ \cite{Dai,Bull}, one
can calculate the expected temperature dependence of the CEF
levels assuming either $\Gamma_5$ \cite{Dai} or $\Gamma_3$
\cite{Bull} as the zero-energy ground state (Fig. 2c). The
comparison of Figs. 2b and 2c reveals that both CEF models are
incompatible with the data.

If excitations in the NFL $x=0.35$ are indeed nonlocalized, one
would expect to find magnetic scattering at energies much less
than 3 meV. Figure 3a shows energy scans at $q=1.3$ \AA$^{-1}$ for
$x=0.35$ and $x=0.0$ using SPINS at
NCNR. Consistent with results at higher energies (Figs. 1 and 2),
magnetic excitations between 0.4 meV and 8 meV are broad,
featureless, and temperature independent from 1.4 K to 300 K. To
see if magnetic scattering in $x=0.35$ peaks at the same AF wave
vector as Y$_{0.55}$U$_{0.45}$Pd$_3$ \cite{Dai}, we carried out a
series of energy scans at different wave vectors at $T=1.4$ K. The
outcome shows no enhancement along any wave vectors probed (Fig.
3b). To see if there is magnetic scattering at an arbitrary
($q=1.95$ \AA$^{-1}$) elastic position, we performed polarized
neutron beam measurements on $x=0.35$ at $T=5$ K using BT-2 at
NCNR. The flipping ratios for both horizontal and vertical guide
fields are $\sim$20. By subtracting vertical field intensity from
that in horizontal field, we confirmed the presence of elastic
magnetic scattering \cite{Dai,note}. To further prove that the
observed excitations are from U moments, we show in Fig. 3c the
wave vector dependence of the magnetic scattering from both HET
and SPINS experiments normalized to the expected U$^{4+}$ and
U$^{3+}$ magnetic form factors. The data are clearly consistent
with U magnetic scattering.

The absence of any characteristic $q$ and $E$ scale in the
magnetic excitations of the $x=0.35$ compound suggests that
isolated U ions are responsible for the observed spin dynamical
behavior. The unique temperature independent form of the magnetic
scattering $S(q,\omega,T)$ bears a remarkable resemblance to that
of UCu$_{5-x}$Pd$_x$, where the excitations at all $q$, and for
limited temperatures and energies ($<25$ meV) accessed display the same type of
NFL $\omega/T$ scaling \cite{Aronson,Aronson2}. The
measured $S(q,\omega,T)$ is related to the imaginary part of the
dynamical susceptibility, $\chi^{\prime\prime}(q,\omega,T)$, via
$S(q,\omega,T)=\chi^{\prime\prime}(q,\omega,T)/[1-\exp(-\hbar\omega/k_{B}T)]$,
where $[n(\omega) +1 ]=1/[1-\exp(-\hbar\omega/k_{B}T)]$ is the
Bose population factor. In calculating
$\chi^{\prime\prime}(q,\omega,T)$ for the various temperatures, we
find that $\chi^{\prime\prime}$ multiplied by $T^{1/5}$ collapses
onto a single curve for all data sets as a function of
$\omega$/$\textit{T}$.

Figure 4 shows the outcome of our analysis, where the SPINS data
have been scaled to the absolute scale of the HET data through
normalizing the elastic incoherent scattering of $x=0.0$ and 0.35.
In the final plot, all data have been corrected for their magnetic
form factor dependence, which is critical for the time-of-flight
data because of the coupled $E$-$q$ values. The obtained
scaling exponent of 1/5 represents a purely empirical analysis;
however, slight deviations from this value induce substantial
discontinuities in the resulting $\omega/{T}$ scaling plot. For comparison,
the scaling exponent of UCu$_{5-x}$Pd$_x$ is 1/3 \cite{Aronson,Aronson2}.

The discovery of $\omega/T$ scaling in the NFL $x=0.35$ compound
strongly suggests that the magnetic fluctuations in this system
arise from the close proximity to a $T = 0$ K phase
transition. Similar $\omega/T$ scaling was first identified in the
NFL UCu$_{5-x}$Pd$_x$ system, but with much smaller energy range \cite{Aronson,Aronson2}. The key
difference, however, is that Sc$_{0.65}$U$_{0.35}$Pd$_3$ does not have
any enhancement in the magnetic scattering around the expected AF ordering
vector of higher U concentrations \cite{Dai}. While the antiferromagnetism in
Y$_{1-x}$U$_x$Pd$_3$ compounds with $x\geq 0.41$ may not control the spin dynamics for Sc$_{0.65}$U$_{0.35}$Pd$_3$,
our results are consistent with the observation that the
spin-glass transition temperatures of the NFL $x=0.35$ and 0.3
compounds are suppressed close to $T=0$ K \cite{Dickey}. Since NFL
behavior has previously been attributed to the proximity of an AF
QCP at $T=0$ K in CeCu$_{5.9}$Au$_{0.1}$ \cite{Schroeder}, our
results suggest that details of the $T=0$ K phase transition are
unimportant for the NFL behavior.

\begin{figure}[t]
\includegraphics[scale=.4]{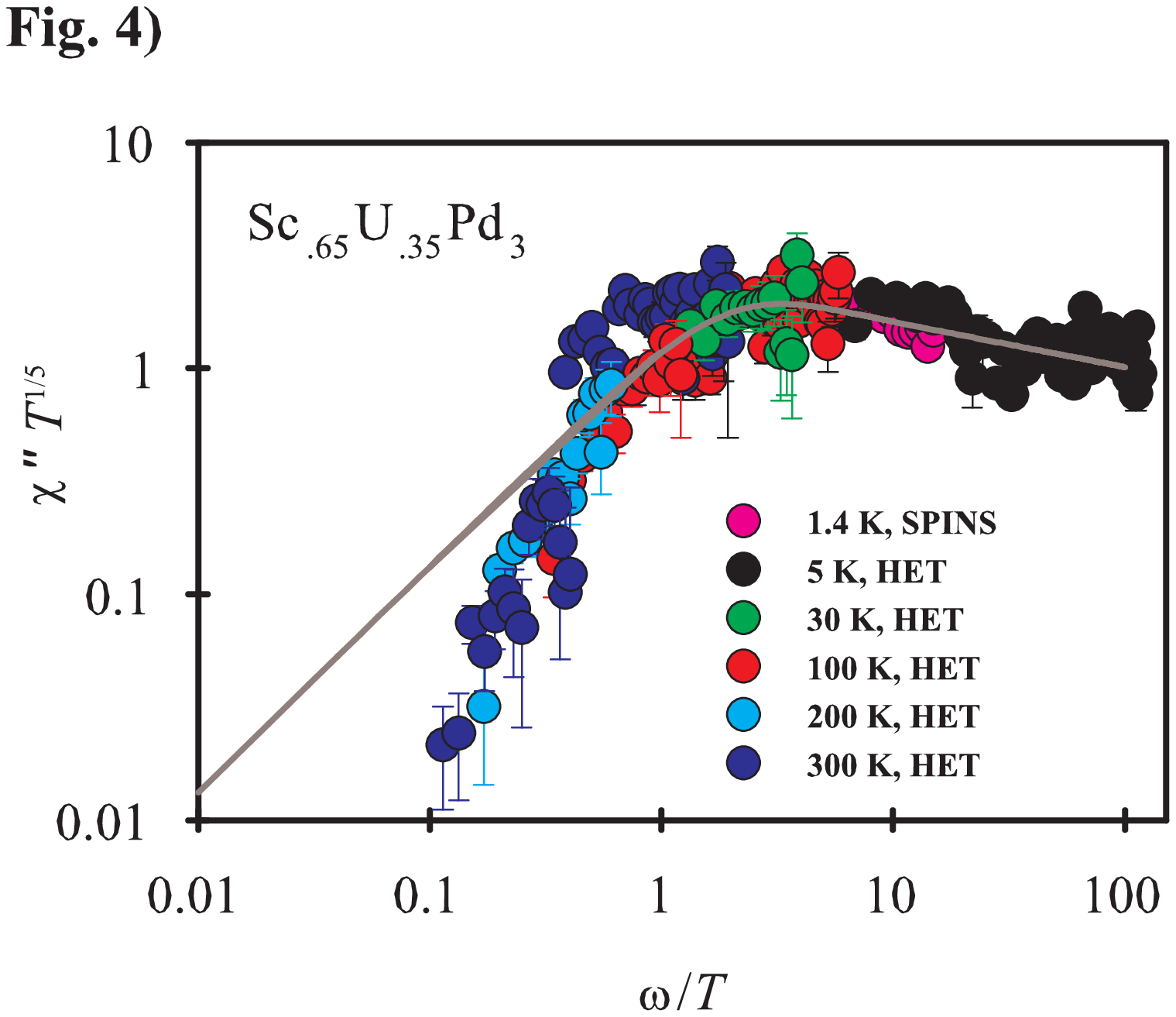}
\caption{Scaling plot for Sc$_{.65}$U$_{.35}$Pd$_3$. The 300K HET
data exhibit a slight deviation due possibly to underestimation
of phonon contributions.}
\end{figure}

Theoretically, the NFL behavior may arise from the proximity to a
$T=0$ K spin-glass quantum phase transition, although models in
their present forms do not predict the observed $\omega/T$ scaling
\cite{Sachdev, Sengupta,Georges}. Recent experiments on
Ce(Ru$_{0.5}$Rh$_{0.5}$)$_{2}$Si$_{2}$ \cite{Tabata} have
attributed the NFL behavior to the disorder near a spin glass QCP
\cite{Miranda,Neto,Grempel}. On the other hand,
disorder was found not to be the main cause for the NFL behavior
in quantum spin-glasses UCu$_{5-x}$Pd$_x$ at $x=1.0$ and 1.5
\cite{MacLaughlin,Bauer}.
Assuming that disorder does not play a major role
\cite{Gajewski,Dickey}, one can envision three different
microscopic scenarios for the NFL behavior in
(Y,Sc)$_{1-x}$U$_{x}$Pd$_3$. The first is
the QKE \cite{Cox}, where one would expect localized spin
excitations with nonmagnetic $\Gamma_3$ as the ground state.
Inspection of previous data for Y$_{0.8}$U$_{0.2}$Pd$_3$
\cite{Dai, Bull} as well as Figs. 1-3 for
Sc$_{0.65}$U$_{0.35}$Pd$_3$ reveals no convincing evidence for
localized states. In addition, there is clear magnetic scattering
at $E=0$ meV, and the temperature dependence of magnetic
excitations does not follow the expectations of a simple CEF
scheme (Figs. 1-3). The second is the $T=0$ K AF phase transition
\cite{Andraka,Si}. However, $\chi^{\prime\prime}(q,\omega,T)$
displays localized moment dynamics with no evidence for U-U
correlations (Fig. 1a) \cite{noter}. Instead, the data are consistent with
$\omega/T$ scaling analogous to UCu$_{5-x}$Pd$_x$, and therefore
can be understood as manifestations of single-impurity critical
scaling associated with a spin-glass phase transition suppressed
to near 0 K \cite{Aronson3}. The solid line in Fig. 4
shows the theoretically proposed spin susceptibility scaling
function
$\chi^{\prime\prime}(q,\omega,T)=1/[AT^{\alpha}F(\omega/T)]$ with
$\alpha=1/5$ and $F(\omega/T)=\exp[\alpha\Psi(1/2-i\omega/2\pi
T)]$ \cite{Si,note2}. Although notable deviations with the opposite sign
from UCu$_{5-x}$Pd$_x$ are seen for small $\omega/T$ \cite{Aronson2}, the model accurately
describes the data over a remarkable $\omega/T$ range (Fig. 4).

In summary, we have used inelastic neutron scattering to show that
magnetic excitations in the NFL Sc$_{1-x}$U$_{x}$Pd$_3$ ($x=0.35$)
compound are broad and featureless in wave vector and energy. The
absence of any characteristic energy scale, other than the
temperature itself, suggests that the microscopic origin of the
NFL behavior lies with individual U
ions near a $T=0$ K spin-glass phase transition. Therefore, the
NFL properties in a wide variety of $f$-electron systems including
(Y,Sc)$_{1-x}$U$_{x}$Pd$_3$,
UCu$_{5-x}$Pd$_x$, CeCu$_{6-x}$Au$_x$, and Ce(Rh$_{0.8}$Pd$_{0.2}$)Sb can be described by a
common physical picture, being near a $T=0$ K quantum phase
transition. Although the intrinsic disorder in these systems
is essential for establishing the spin-glass ground state \cite{Miranda,Neto,Grempel}, it cannot be the main
cause of the NFL behavior.

We thank M. C. Aronson, H. J. Kang, and Q. Si for helpful discussions.
This work is supported by the US NSF DMR-0139882 and DOE DE-AC05-00OR22725 at UT/ORNL, and by the US DOE DE-FG02-04ER46105 and NSF DMR-0335173 at UCSD.


\begin{thebibliography}{}
\bibitem{note} Intensities
for horizontal and vertical fields are $679.7\pm 8.2$ and
$655.7\pm 7.4$ counts/hour, respectively. The net elastic magnetic
scattering for $x=0.35$ is then $48\pm 22$ counts/hour at $T=5$ K
and $q=1.95$ \AA $^{-1}$.
\bibitem{noter} The weak $q$-peak in Fig. 10a of Ref. \cite{Bull}
is then due to the U-U clustering in Y$_{0.8}$U$_{0.2}$Pd$_3$ \cite{Sullow}.
\bibitem{note2} We use this functional form although it was derived for a AF QCP \cite{Si}, not
a spin-glass QCP.
\end{thebibliography}

\end{document}